\documentclass[preprint2]{aastex}

\newcommand{\sax}{{\it BeppoSAX\,\,}}
\newcommand{\lae}{\mathrel{<\kern-1.0em\lower0.9ex\hbox{$\sim$}}}
\newcommand{\gae}{\mathrel{>\kern-1.0em\lower0.9ex\hbox{$\sim$}}}
\newcommand{\ecs}{{\rm erg} \, {\rm cm}^{-2} \, {\rm s}^{-1}} 
\newcommand{\keV}{\,{\rm keV} }
\newcommand{\MeV}{\,{\rm MeV} }
\shorttitle{First INTEGRAL AGN catalog}
\shortauthors{Beckmann et al.}

\begin{document}

\title{The First INTEGRAL AGN Catalog}


\author{V. Beckmann\altaffilmark{1}, N. Gehrels, C. R. Shrader\altaffilmark{2}}
\affil{NASA Goddard Space Flight Center, Exploration of the Universe Division, Code 661, Greenbelt, MD 20771, USA}
\email{beckmann@milkyway.gsfc.nasa.gov}
\author{S. Soldi\altaffilmark{3}}
\affil{INTEGRAL Science Data Centre, Chemin d' \'Ecogia 16, 1290
 Versoix, Switzerland}


\altaffiltext{1}{also with the Joint Center for Astrophysics, Department of Physics, University of Maryland, Baltimore County, MD 21250, USA}
\altaffiltext{2}{also with Universities Space Research Association, 10211 Wincopin Circle, Columbia, MD 21044, USA}
\altaffiltext{3}{also with Observatoire de Gen\`eve, 51 Ch. des Maillettes, 1290 Sauverny, Switzerland}

\begin{abstract}
We present the first {\it INTEGRAL} AGN catalog, based on observations performed from launch of the mission in October 2002 until January 2004. The catalog includes 42 AGN, of which 10 are Seyfert~1, 17 are Seyfert~2, and 9 are intermediate Seyfert~1.5. The fraction of blazars is rather small with 5 detected objects, and only one galaxy cluster and no star-burst galaxies have been detected so far. A complete subset consists of 32 AGN with a significance limit of $7 \sigma$ in the {\it INTEGRAL}/ISGRI 20--40 keV data.
Although the sample is not flux limited, the distribution of sources
shows a ratio of obscured to unobscured AGN of $1.5 - 2.0$, consistent
with luminosity dependent unified models for AGN. Only four Compton-thick AGN are found in the sample. 
Based on the {\it INTEGRAL} data presented here, the
Seyfert~2 spectra are slightly harder ($\Gamma = 1.95 \pm 0.01$) than
Seyfert~1.5 ($\Gamma = 2.10 \pm 0.02$) and Seyfert~1 ($\Gamma = 2.11
\pm 0.05$).  
\end{abstract}




\section{Introduction}
The X-ray sky as seen by X-ray satellites over the past 40 years, shows a 
substantially different picture than for example the optical band. While the visual 
night sky is dominated by main sequence stars, Galactic binary systems and super 
nova remnants form the brightest objects in the X-rays. Common to both regimes 
is the dominance of active galactic nuclei (AGN) toward lower fluxes. 
In the X-ray range itself, one observes a slightly different
population of AGN at soft and at hard X-rays. Below 5 keV the X-ray
sky is dominated by AGN of the Seyfert 1 type, above the absorbed
Seyfert 2 objects appear to become more numerous. 
These type 2 AGN are also the main contributors to the cosmic X-ray
background above 5 keV \citep{Setti,Comastri1995,HELLAS,hardXRB}, although 
only $\sim 50 \%$ of the XRB above $8 \rm \, keV$ can be resolved \cite{Worsley}.
Beside the rather persistent Seyfert galaxies, also blazars are detectable 
in the hard X-rays. Because of the fact that we look into a highly relativistic 
jet in the case of those sources, the blazars exhibit strong variability 
on all time-scales and are especially variable in the X-ray and high-energy gamma-ray domain. 
For an overview over the various extragalactic X-ray surveys see Brandt \& Hasinger (2005),
and for early {\it INTEGRAL} results see Krivonos et al. (2005). 

The hard X-ray energy range is not currently accessible to X-ray telescopes using grazing 
incidence mirror systems. Instead detectors without spatial resolution like the 
PDS \cite{pds} on {\it BeppoSAX} (Boella et al. 1997) and OSSE \cite{OSSE} on {\it CGRO} (Gehrels et al. 1993) 
have been applied. 
A synopsis of these previous results is as follows: the 2 -- 10 keV Seyfert~1 continua 
are approximated by a $\Gamma \simeq 1.9$ powerlaw form \cite{Sy1average}. A flattening 
above $\sim 10 \keV$ has been noted, and is commonly attributed to Compton reflection \cite{george}. 
There is a great deal of additional detail in this spectral domain - ``warm'' absorption, 
multiple-velocity component outflows, and relativistic line broadening - which are beyond the 
scope of this paper. The Seyfert~2 objects are more poorly categorized here, but the general 
belief is that they are intrinsically equivalent to the Seyfert~1s, but viewed through much 
larger absorption columns [see e.g. V\'eron-Cetty \& V\'eron (2000) 
for a recent review]. 

Above 20 keV the empirical picture is less clear. The $\sim 20 - 200 \keV$ continuum 
shape of both Seyfert types is consistent with a thermal Comptonization spectral form, 
although in all but a few cases the data are not sufficiently constraining to rule out 
a pure powerlaw form. Nonetheless, the non-thermal scenarios with pure powerlaw continua 
extending to $\sim \MeV$ energies reported in the pre-{\it CGRO} era are no longer 
widely believed, and are likely a result of background systematics. However, a detailed 
picture of the Comptonizing plasma - its spatial, dynamical, and thermo-dynamic 
structure - is not known. Among The critical determinations which {\it INTEGRAL} or future hard 
X-ray instruments will hopefully provide are the plasma temperature and optical depth 
(or Compton ``Y'' parameter) for a large sample of objects. 

The other major class of gamma-ray emitting AGN - the blazars (FSRQs and BL Lac objects) are 
even more poorly constrained in the {\it INTEGRAL} spectral domain. The detection of $\sim 70$ 
of theses objects with the EGRET instrument on {\it CGRO} in the 1990s has established the 
presence of $\Gamma \sim 10$ relativistic plasma at small angles to the observers line 
of sight. The so called ``TeV'' blazars are similar objects but with the peak power 
shifted to even greater energies, probably due to an even smaller viewing angle (see for
example, Fossati et al. 1999, and for earlier {\it INTEGRAL} results
Pian et al. 2005).  The most 
common interpretation of the broad-band continuum emission is the synchrotron self Compton 
(SSC) model, although the external Compton models (EC) are still also plausible. Broad-band 
temporal coverage can in principle discriminate between these possibilities,
although each  of these components can separately contribute to the hard-X-ray band under
extreme circumstances. The {\it INTEGRAL} 
bandpass covers a critical region which can in principle establish the amplitude and shape 
of the high-end synchrotron component or or low-end SSC component. 

Critical to each of these issues is the need to obtain improved continuum measurements over 
the hard X-ray to soft gamma-ray range for as large a sample of objects as possible. 
{\it INTEGRAL} \cite{INTEGRAL}, since its launch in October 2002, offers unprecedented 
$> 20 \keV$ collecting area and state of the art detector electronics and background 
rejection capabilities. Thus it offers hope of substantial gains in our knowledge of 
the AGN phenomenon. 
We were thus motivated at this juncture some three years into the 
mission, to formulated and present the collective look at the data so far obtained and 
assess the potential for progress in the {\it INTEGRAL} era.

As the satellite spent most of its observing time up to now concentrating on the Galactic 
plane, some low-latitude AGN have been detected. Observations at higher Galactic latitude 
have also led to studies of individual sources: 
3C~273 \cite{3c273}, NGC~4388 \cite{NGC4388}, \mbox{GRS~1734--292} (Sazonov et al. 2004),
the S5~0716+714 field \cite{0716}, \mbox{PKS~1830--211} \cite{PKS1830}, NGC~4151 (Beckmann et al. 2005), 
six AGN near the Galactic plane (Soldi et al. 2005), and an early study of all extragalactic 
sources seen through the Galactic plane \cite{Bassani}. Two of the new sources discovered 
by {\it INTEGRAL}, \mbox{IGR~J18027-1455} and IGR~J21247+5058, have also been shown to 
be AGN \cite{IGRsources}. 

With the on-going observation of the sky by {\it INTEGRAL}, a sufficient amount of data is now accessible to compile the first {\it INTEGRAL} catalog of AGN. 
In this paper we present analysis of recent observations performed by the {\it INTEGRAL} satellite instruments IBIS/ISGRI, SPI, and JEM-X, and compare the results with previous studies.

\section{Observations}

Observations in the X-ray to soft gamma-ray domain have been performed by the instruments on-board the {\it INTEGRAL} satellite \cite{INTEGRAL}. This mission offers the unique possibility to perform simultaneous observations over the $2 - 8000 \rm \, keV$ energy region. This is achieved by the X-ray monitor (2--30 keV) JEM-X \cite{JEMX}, the soft gamma-ray imager (20--1000 keV) ISGRI \cite{ISGRI}, and the spectrograph SPI \cite{SPI}, which operates in the 20 -- 8000 keV region. Each of these instruments employs the coded-aperture technique \cite{codedmask}. 
In addition to these data an optical monitor (OMC, Mas-Hesse et al. 2003) provides photometric measurements in the $V$ band.  

\subsection{{\it INTEGRAL} Data}
\label{data}

Data used for the analysis presented here were all in the public domain by the end of March 2005. This includes data from orbit revolutions 19 - 137 and revolutions 142 - 149. Data before revolution 19 have been excluded as the instruments settings changed frequently and therefore the data from this period are not suitable to be included into a homogeneous survey. Most of the {\it INTEGRAL} observing time has been spent on the Galactic Plane so far, with some deep exposures at high Galactic latitude, for example in the Virgo and the Centaurus region.

Our sample objects have been selected in three ways. The first criterion was a detection in the ISGRI data in a single pointed observation. The duration of one pointing of the spacecraft is usually 2 - 3 ks long. The standard analysis of the data is performed at the {\it INTEGRAL} Science Data Centre (ISDC; Courvoisier et al. 2003b). These results are publicly available\footnote{\tt http://isdc.unige.ch/index.cgi?Data+sources}. Within one pointing the detection threshold for a source in the center of the field of view of ISGRI is about $20 \rm \, mCrab$. 
A second level search was done using combined data sets of individual observations and contained within separate revolutions which are about 3 days long. In cases where an observation lasted longer than a revolution, the data set was split by revolution. The results from this analysis are publicly available through the HEASARC archive\footnote{\tt http://heasarc.gsfc.nasa.gov/docs/integral/inthp\_archive.html}. Clearly the sensitivity of this second method depends strongly on the duration of the analyzed observation. The sensitivity also depends on the dither pattern of the observation and on the position of the source within the field of view. 
In addition to those two ways of combining the list of AGN seen by {\it INTEGRAL}, we added all AGN which have been reported as {\it INTEGRAL} detections so far in the literature.
This combined list of objects is presented in Table~\ref{catalog}. In order to get meaningful results for each of the sources in this AGN sample, it was necessary to perform an individual analysis for each one. 
We used the data for which the spacecraft was pointed within 10
degrees radius about each AGN for ISGRI and SPI, and within 5 degrees
for JEM-X. The exposure times listed in Tab.~\ref{catalog} refer to the effective ISGRI values.  These values are approximately the same for the spectrograph SPI, but for the JEM-X monitor it has to be taken into account that they cover a much smaller sky area. Thus in the case of dithering observation, the source is not always in the field of view of the monitors, and the exposure times are correspondingly shorter.

The analysis was performed using the Offline Science Analysis (OSA)
software version 5.0 distributed by the ISDC (Courvoisier et
al. 2003b). This version shows substantial improvement compared to the
previous OSA 4.2 version, except for the JEM-X spectral extraction,
for which we used the March 15, 2005 version of the OSA. 
 The significances listed in Tab.~\ref{catalog} have been derived by using the OSA software for ISGRI (20 -- 40 keV), SPI (20 -- 40 keV), and JEM-X (2 -- 20 keV). In cases where there were no publicly available data covering the source position, we put ``n/a'' in the significance column. In those cases where no JEM-X or SPI detection was achieved, ``--'' is written.
For comparison we list the photon flux measured by {\it CGRO}/OSSE in the $15 - 150 \keV$ energy band, as reported by Johnson et al. (1997) for Seyfert galaxies and by McNaron-Brown et al. (1995) for blazars.
Apart from Gamma-ray bursts, no extragalactic source detections by
IBIS/PICsIT have been reported.

The analysis of the {\it INTEGRAL}/IBIS data is based on a cross-correlation procedure between the recorded image on the detector plane and a decoding array derived from the mask pattern \cite{IBISOSA}.
The ISGRI spectra have been extracted from the count rate and variance
mosaic images at the position of the source, which in all cases
corresponds to the brightest pixel in the $20 - 40 \rm \, keV$ band. 
The SPI analysis was done using the specific analysis software
\cite{SPIOSA} including version 9.2 of the reconstruction software
SPIROS \cite{SPIROS} which is based on the ``Iterative Removal of
Sources'' technique (IROS; Hammersley et al. 1992): a simple image of
the field of view is made using a mapping technique which is optimized
for finding a source assuming that the data can be explained by only
that source plus background. The mapping gives the approximate
location and intensity of the source, which are then improved by
maximizing a measure of the goodness of fit. The residuals of the fit
are used as the input for a further image reconstruction and source
search \cite{SPIROS}. The fitting process is easier when an input
catalog is provided. In this work we used the source positions found
with ISGRI as an input catalog for the SPI analysis, as the
sensitivity of ISGRI is higher than that of SPI below 200 keV.

With respect to the calibration uncertainty, the IBIS instrument team stated that the systematic error is of the order of 1.5\% (2005, private communication). Nevertheless, this value corresponds to on-axis observations within a short period of time with no disturbing influence, such as enhanced background activity. A combined fit of Crab spectra taken in revolution 43, 44, 120, 170, and 239, i.e. over a 1.5 year span of the mission, shows a larger uncertainty in the flux. The scale of diversity, assuming that Crab is a source with constant flux, gives some hint what is the scale of
discrepancies in count rates observed in various conditions, i.e.,
with different dithering patterns or instrumental settings. Clearly the 
absolute calibration cannot be more precise than the observed
variations. Hence, it seems that for the 22-120 keV band we may conclude that the joint $1 \sigma$ uncertainty of the spectral extraction method and the calibration files is about $5\%$. It may happen that for other sources and in some special circumstances
we may encounter a larger discrepancy especially for faint sources,
but a systematic error of 5\% should be a valid approximation. 
In most cases the fit to the data from the three instruments
used in this work did not show any improvement when adding variable
cross-calibration. This is mainly due to the fact that most spectra
have low signal-to-noise and adding a further parameter to the fit
procedure worsens the fit result, according to an F-test.

The spectral shape calibration was also tested on {\it INTEGRAL} Crab
observations. With the so-called canonical model for the Crab showing
a single power law with photon index $\Gamma = 2.1$ \cite{Crab}, the
values retrieved from {\it INTEGRAL} are $\Gamma = 2.1$ (JEM-X), and $\Gamma = 2.2$ for SPI and ISGRI \cite{Kirsch05}. It should be kept
in mind though that the Crab is significantly brighter than the
sources discussed here,
and systematic effects might depend on source brightness.


All the extracted images and source results are available in electronic form\footnote{http://heasarc.gsfc.nasa.gov/docs/integral/inthp\_archive.html}.


\subsection{{\it CGRO} Data}

{\it CGRO}/OSSE \cite{OSSE} covered the energy range $50 - 10,000 \rm \, keV$. It therefore preferentially detected AGN with hard and bright X-ray spectra, which we also expect to be detectable by {\it INTEGRAL}. Several of the AGN detected by OSSE have not been seen by {\it INTEGRAL} so far. We list those sources and the {\it INTEGRAL}/ISGRI exposure time on the particular AGN in Table~\ref{OSSEnonINTEGRAL}$^3$.

\section{X-ray to Gamma-ray Spectra}

\subsection{Time Averaged Spectral Analysis}

\begin{figure}
\plotone{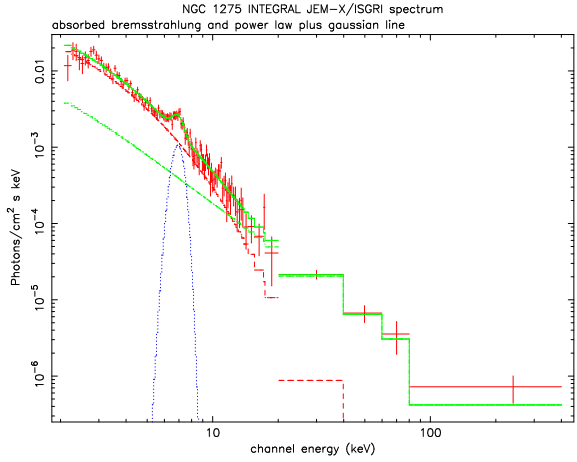}
\caption[]{{\it INTEGRAL} spectrum of \mbox{NGC~1275}.}
\label{fig:NGC1275}
\end{figure}

\begin{figure}
\plotone{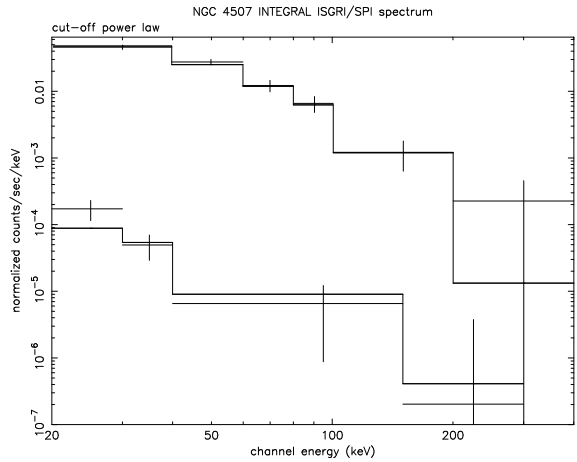}
\caption[]{{\it INTEGRAL} count spectrum of \mbox{NGC~4507}. No JEM-X data are available for this source. We note that here, and elsewhere SPI spectral data are presented, the ``detector counts'' have really been effective area corrected by the image reconstruction algorithm \cite{SPIROS}. Thus, the SPI spectra are really $\rm photons \, cm^{-2} \, s^{-1} \, keV^{-1}$.}      
\label{fig:NGC4507}
\end{figure}

\begin{figure}
\plotone{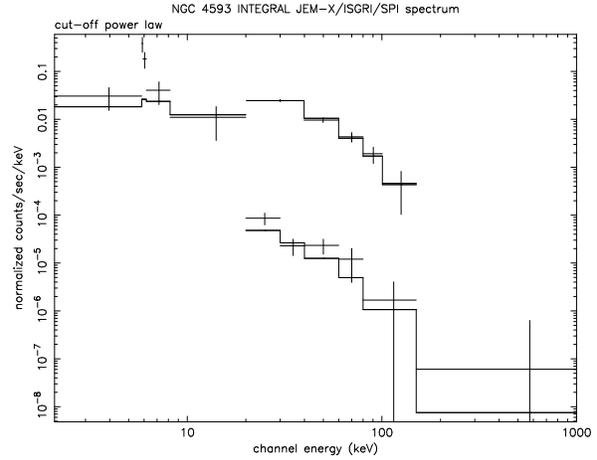}
\caption[]{{\it INTEGRAL} count spectrum of \mbox{NGC~4593}.}      
\label{fig:NGC4593}
\end{figure}

\begin{figure}
\plotone{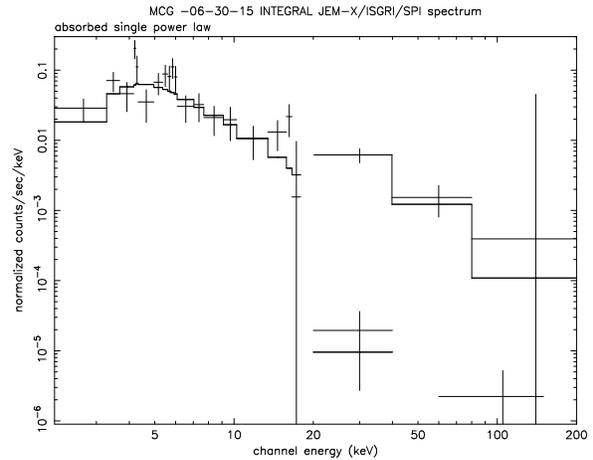}
\caption[]{{\it INTEGRAL} count spectrum of \mbox{MCG~--06--30--015} with JEM-X (2 -- 18 keV), ISGRI (20 -- 200 keV; upper spectrum), and SPI (lower two points).}      
\label{fig:MCG6}
\end{figure}

\begin{figure}
\plotone{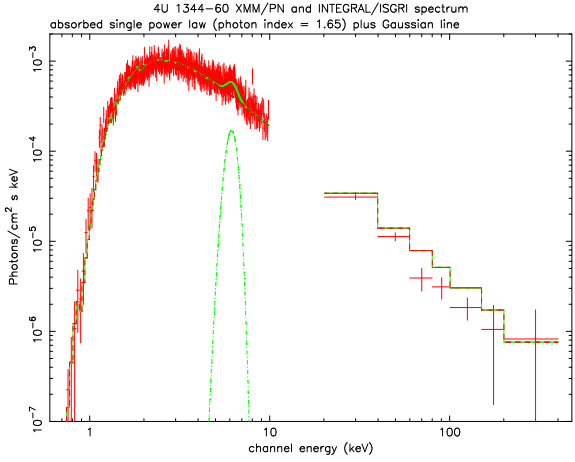}
\caption[]{Combined XMM-Newton EPIC/pn and {\it INTEGRAL} ISGRI
  photon spectrum of \mbox{4U~1344--60}. The SPI spectrum is not shown
  for better readability. The iron K$\alpha$ line indicates a redshift
  of about $z = 0.043$.}      
\label{fig:4U1344}
\end{figure}

\begin{figure}
\plotone{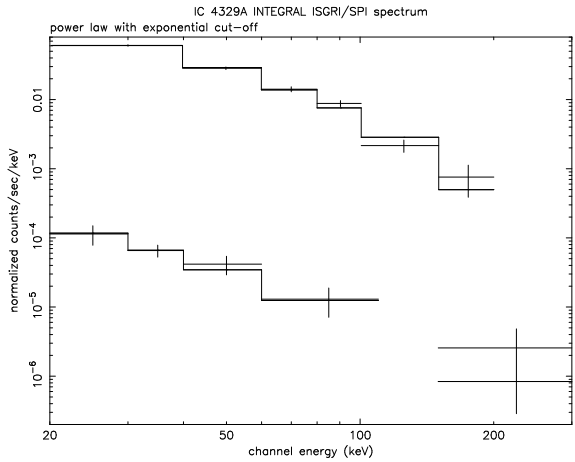}
\caption[]{{\it INTEGRAL} ISGRI and SPI count spectrum of IC~4329A.}      
\label{fig:IC4329A}
\end{figure}

\begin{figure}
\plotone{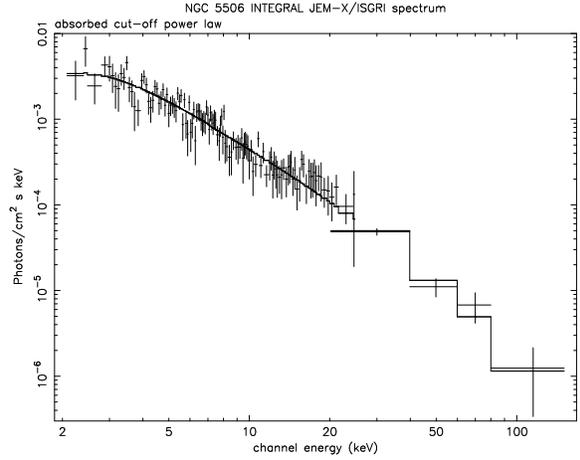}
\caption[]{{\it INTEGRAL} spectrum of NGC~5506.}      
\label{fig:NGC5506}
\end{figure}

\begin{figure}
\plotone{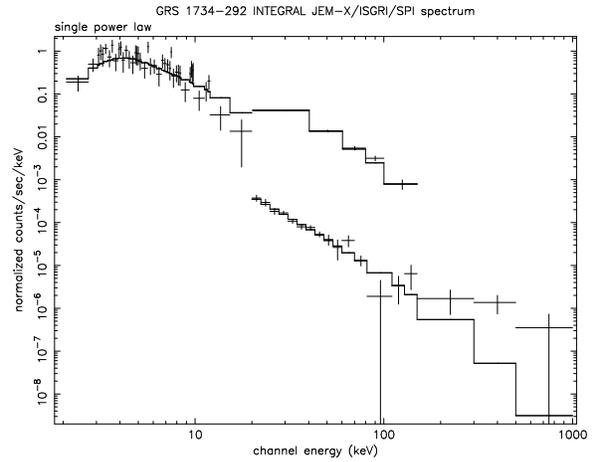}
\caption[]{{\it INTEGRAL} count spectrum of GRS~1734--292.}      
\label{fig:GRS1734}
\end{figure}

\begin{figure}
\plotone{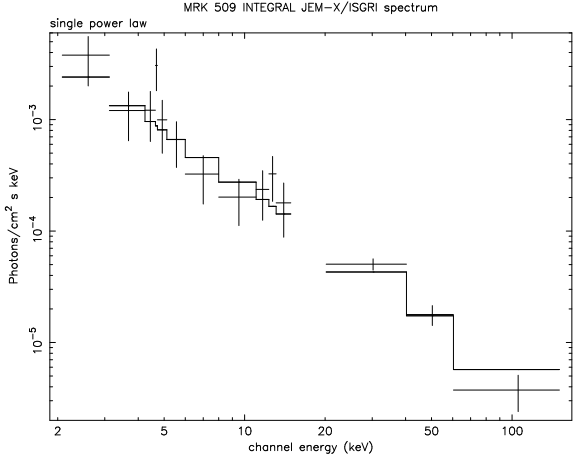}
\caption[]{{\it INTEGRAL} spectrum of MRK~509. No SPI data are available for this source.}      
\label{fig:MRK509}
\end{figure}

\begin{figure}
\plotone{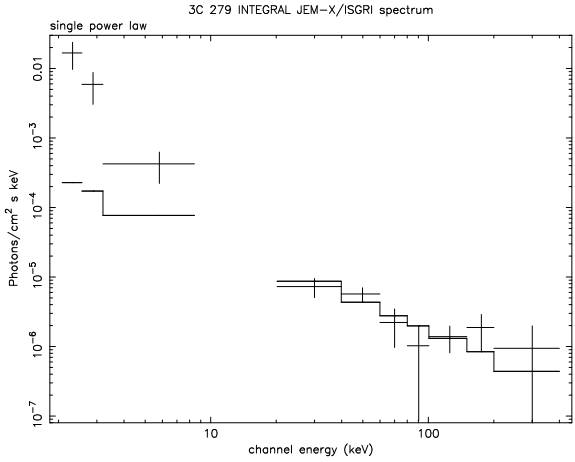}
\caption[]{{\it INTEGRAL} spectrum of the blazar \mbox{3C~279}. The source is not detectable in the SPI data.}      
\label{fig:3C279}
\end{figure}

\begin{figure}
\epsscale{0.7}
\plotone{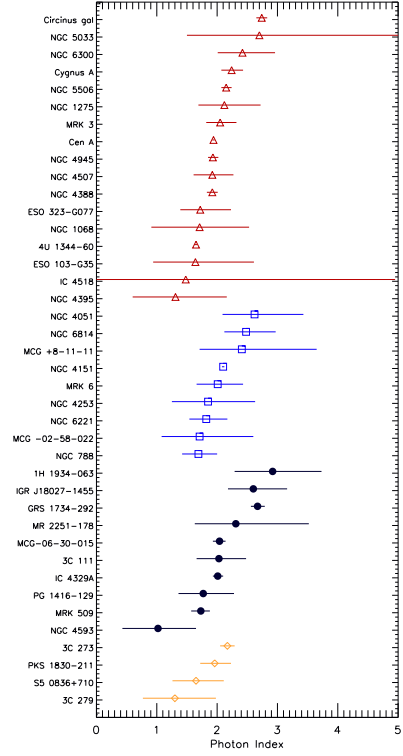}
\caption[]{Photon index of a single power law fit to the {\it
    INTEGRAL} data of our sample, ordered in four
  categories. From top to bottom: Seyfert~2, Seyfert~1.5, Seyfert~1,
  and blazars.}      
\label{fig:photonindices}
\end{figure}

The overall spectra extracted from the {\it INTEGRAL} JEM-X, ISGRI and
SPI data for 10 AGNs are shown in Fig.~\ref{fig:NGC1275} to
Fig.~\ref{fig:3C279} (see Sect.~4 for details). A single powerlaw fit
with absorption from material in the line of sight has been
applied to the {\it INTEGRAL} spectra of all sources and the results
are presented in Table~\ref{fitresults} and in
Figure~\ref{fig:photonindices}. The errors on the photon indices are
90\% confidence errors. The intrinsic absorption was
determined by subtracting the Galactic hydrogen column density from
the total absorption as measured in the soft X-rays. In the cases
where the {\it INTEGRAL}/JEM-X data were not available or did not
restrict the absorption column density sufficiently,
values from the literature have been used to fix this component in the
spectral fit (Tab.~\ref{fitresults}). No systematic error has been added to the data.
The spectral fitting was done using version 11.3.2 of XSPEC
\cite{XSPEC}. In most cases a single powerlaw gave an acceptable fit
to the data, which also is based on the fact that the signal-to-noise
of many of the spectral data is rather low. In six cases a cut-off
power law fit led to better fits results. We report those results in
Table~\ref{cutoff}.

\subsection{Averaged Spectra of AGN Subtypes}

In order to investigate the spectra of AGN subtypes, we have derived
averaged spectra of the Seyfert 1 and 2 types, as well as for the
intermediate Seyferts and the blazars, and according to the intrinsic
absorption. 
The average Seyfert 1 spectrum was constructed using the weighted mean
of 10 ISGRI spectra, the Seyfert 2 composite spectrum includes 15
sources, and 8 objects form the intermediate Seyfert 1.5 group. The
two brightest sources, Cen~A and NGC~4151, have been excluded from the
analysis as their high signal-to-noise ratio would dominate the
averaged spectra.
The average spectra have been constructed by computing the weighted mean of all fit results on the individual sources as shown in
Table~\ref{fitresults}. In order to do so, all spectra had been fit by an absorbed single
powerlaw model. When computing the weighted average of the various
sub classes, the Seyfert 2 objects show flatter hard X-ray spectra
($\Gamma = 1.95 \pm 0.01$) than the Seyfert 1.5 ($\Gamma = 2.10 \pm
0.02$), and Seyfert 1 appear to have the steepest spectra ($\Gamma =
2.11 \pm 0.05$) together with the blazars ($\Gamma = 2.07 \pm 0.10$). 

The classification according to the Seyfert type of the objects is
based on optical observations. An approach to classify sources due to
their properties in the X-rays can be done by separating the sources
with high intrinsic absorption ($N_{\rm H} > 10^{22} \rm \, cm^{-2}$)
from those objects which do not show significant absorption in the
soft X-rays. It has to be pointed out that not all objects which show
high intrinsic absorption in the X-rays are classified as Seyfert~2
galaxies in the optical, and the same applies for the other AGN
sub-types. Nevertheless a similar trend in the spectral slopes can be
seen: the 21 absorbed AGN show a flatter hard X-ray spectrum ($\Gamma
=1.98 \pm 0.01$) than the 13 unabsorbed sources ($\Gamma
=2.08 \pm 0.02$). The blazars have again been excluded from these
samples. 

\section{Notes on Individual Sources}

Spectra are presented in this paper for {\it INTEGRAL} sources which are as yet not published and for which at least two instruments yield a detection of $> 3\sigma$. Spectral fits did not require any cross-calibration correction between the instruments, except for GRS~1734-292 and for the two brightest sources, Cen~A and NGC~4151. In cases where the spectra of different instruments overlap, the spectral plots have been performed in count space, otherwise in photon space.

{\bf NGC 1275}, a Seyfert 2 galaxy at $z = 0.018$, was detected in the
ISGRI and JEM-X data (Fig.~\ref{fig:NGC1275}). The spectrum represents
the blend of the Perseus cluster with the spectrum of the AGN. The
measured spectrum in the $2 - 400 \rm \, keV$ energy range is fitted
by a slightly absorbed ($N_{\rm H} = (6.7 {+8.1 \atop -6.7}) \times
10^{21} \rm \, cm^{-2}$) bremsstrahlung model ($kT = 3.4 {+0.4 \atop
  -0.4}$) plus a powerlaw with a photon index fixed to the measurement
from the ISGRI data ($\Gamma = 2.12$). In addition a Gaussian line
with $E = 6.9 { +0.2  \atop -0.2 } \rm \, keV$ and an equivalent width of $EW = 670 \rm \, eV$ has been applied to achieve a reasonable fit result ($\chi^2_\nu = 0.98$). We see no evidence of flattening of the continuum above $\sim 10 \keV$, which is often attributed to Compton reflection, nor can we quantitatively identify a spectral break. 

{\bf NGC~4507} (Seyfert 2; $z = 0.012$) was detected by ISGRI and SPI (Fig.~\ref{fig:NGC4507}). The combined data are best represented by a power-law ($\Gamma = 1.0 {+1.7 \atop -0.8}$) with a high energy cut-off ($E_{\rm cut} = 55 {+123 \atop -11} \rm \, keV$). This is consistent with results from {\it Ginga} and OSSE, which showed $\Gamma = 1.3 \pm 0.2$ and $E_{\rm cut} = 73 {+48 \atop -24} \rm \, keV$ \cite{NGC4507OSSE}.

{\bf NGC 4593}, a Seyfert 1 galaxy in the Virgo region at $z =  0.009$, shows a cut-off powerlaw with $\Gamma = 1.0 {+0.6 \atop -0.6}$ and $E_{\rm cut} = 35 {+49 \atop -12} \keV$ and is detected up to 150 keV (Fig.~\ref{fig:NGC4593}) with a flux of $f_{20 -100 \keV} = 6.8 \times 10^{-11} \ecs$ ($\chi^2_\nu = 1.00$). This is significantly different from the spectrum measured by \sax: $\Gamma = 1.94 {+0.06 \atop -0.05}$ and $E_{\rm cut} \gg 222 \keV$ \cite{NGC4593SAX} with a similar flux ($f_{20 -100 \keV} = 7 \times 10^{-11} \ecs$). 

{\bf Cen A}, a Seyfert 2 ($z = 0.0018$) is one of the brightest objects in the sample. The statistics allow in this case to determine the intercalibration factors between JEM-X, ISGRI, and SPI ($1.0 : 0.92{+0.04 \atop -0.04} : 1.10{+0.06 \atop -0.06}$). An absorbed power law plus a Gaussian line gives the best fit  to the data with $N_H = 14.6 {+1.4 \atop -1.4} \times 10^{22} \rm \, cm^{-2}$, $\Gamma = 1.96 {+0.02 \atop -0.02}$, and an equivalent width of the iron K$\alpha$ line of $EW = 108 \rm \, eV$. Adding an exponential cut-off does not improve the fit ($E_C = 870 {+3810 \atop -410} \keV$). The results are consistent with results published on a subset of {\it INTEGRAL} data and with {\it BeppoSAX} data \cite{Soldi}.

{\bf MCG~--06--30--015}: The Seyfert 1 galaxy ($z = 0.0077$) was detected by all three instruments (Fig.~\ref{fig:MCG6}). The spectrum is well represented ($\chi^2_\nu = 1.19$) by a power law with $\Gamma = 2.8 {+0.4 \atop -0.3}$ and weak evidence for absorption of $N_{\rm H} = 5.9 {+8.6 \atop -5.5} \times 10^{22} \rm \, cm^{-2}$ ($\chi^2_\nu = 1.11$).

{\bf 4U~1344--60} is a bright ($f_{20-100 \, \rm keV} = 5.7 \times
10^{-11} \ecs$) X-ray source, which has been unidentified until
Michel et al. (2004) reported the identification of this source as
a low redshift AGN but no further information was provided.
In order to clarify the nature of
this object, we analyzed 31 ks of XMM-Newton data of the Centaurus B field,
taken in August 2001. Although 4U~1344--60 appears to be near to the
rim of the EPIC camera, a spectrum can be extracted from EPIC/pn
data which shows an iron line at $6.13 {+0.08 \atop -0.09} \keV$ with
a flux of $f_{K\alpha} = 1.8 {+0.2 \atop -0.3} \times 10^{-4} \rm \,
ph \, cm^{-2} \, s^{-1}$ and an
equivalent width of $410 \rm \, eV$. If we consider this line to be
the iron K$\alpha$ fluorescence line at 6.4 keV, the redshift of
4U~1344--60 is $z = 0.043 {+0.016 \atop -0.014}$. The simultaneous fit
with the {\it INTEGRAL} ISGRI and SPI data shows that the spectrum is
well represented by an absorbed power law plus Gaussian line with
$N_{\rm H} = 2.64 {+0.07 \atop -0.07} \times 10^{22} \rm \, cm^{-2}$
and $\Gamma = 1.65 {+0.02 \atop -0.03}$ ($\chi^2_\nu = 1.06$ for 711
dof; Fig.~\ref{fig:4U1344}).

{\bf IC~4329A} (Seyfert 1.2; $z = 0.016$) is well represented by 
a powerlaw ($\Gamma = 1.5 {+0.4 \atop -0.4}$) with exponential cut-off ($E_C = 104 {+344 \atop -48} \keV$), see Fig.~\ref{fig:IC4329A}. The results are consistent on a $2 - 3 \sigma$ level with the spectrum measured by \sax ($\Gamma = 1.90 \pm 0.05$, $E_C \simeq 300 \keV$; Perola et al. 2002). 

\mbox{\bf NGC~5506}: The JEM-X and ISGRI spectrum of this Seyfert 1.9 ($z =  0.006$) shows an absorbed ($N_{\rm H} =  2.7 {+2.4 \atop -2.3} \times 10^{22} \rm \, cm^{-2}$) high-energy cut-off spectrum with $\Gamma = 1.81 {+0.25 \atop -0.28}$ and $E_{\rm cut} =  57 {+139 \atop -27} \keV$ ($\chi^2_\nu = 1.07$; Fig.~\ref{fig:NGC5506}). 
No SPI data are available for this source, because SPI was undergoing an annealing cycle during the observations of NGC~5506. The results are consistent with {\it BeppoSAX} measurements \cite{ngc5506}, even though the iron K$\alpha$ line and the reflection component are not detectable in the {\it INTEGRAL} data.

{\bf IC~4518} (VV~780): This Seyfert 2 galaxy ($z = 0.016$) was
reported to be detectable in a 700 ks observation in January 2004 on SN 1006 \cite{IC4518}. 
The source was detected with $8.7 \sigma$ in the $20 - 40 \rm \, keV$
band with a photon index of $\Gamma = 2.2 \pm 0.4$ and a flux of
$f_{20-40 \rm \, keV} = 6.8 \times 10^{-12} \ecs$. Note that in
Tab.~\ref{catalog} and Tab.~\ref{fitresults} we refer to our
measurements, which are consistent with the values derived by Kalemci
et al. 2005.

{\bf GRS~1734--292} (AX~J1737.4--2907): This Seyfert 1 galaxy ($z = 0.021$) is located near to the Galactic Center ($l = 358.9^\circ$, $b = 1.4^\circ$). Because this region is monitored regularly by {\it INTEGRAL}, 3.3 Msec of observations are available in the public data. Even though an ISGRI spectrum of this source has been published already \cite{GRS1734}, we discuss this object in more detail, because Sazonov et al. included neither the JEM-X, nor the SPI data in their study. The fact that the source is located in a crowded region, makes the analysis more difficult, especially for SPI data with its $\sim 2.5^\circ$ spatial resolution. Figure~\ref{fig:GRS1734} shows the combined JEM-X2, ISGRI, and SPI spectrum. The data are represented by a simple powerlaw with $\Gamma = 2.63 {+0.09 \atop -0.09}$ and in this case intercalibration factors had to be applied to achieve an acceptable fit result ($\chi^2_\nu = 1.09$, 161 d.o.f.). Adding a cut-off or an absorption component to the model does not improve the fit. The average flux is $f_{20-40 \rm \, keV} = 9 \times 10^{-11} \ecs$. This source is of additional interest, as it lies within the error circle for the EGRET source 3EG~J1736-2908 \cite{GRS1734}. If this is a real association it would be the only known case of a ``gamma-ray loud'' Seyfert 1.

{\bf MRK 509} (Seyfert 1; $z = 0.034$) was not detected by SPI and yielded only low signal-to-noise spectra with JEM-X and ISGRI (Fig.~\ref{fig:MRK509}).
No absorption component was necessary to fit the combined spectrum. A pure single powerlaw ($\Gamma = 1.66 {+0.15 \atop -0.16}$) resulted in $\chi^2_\nu = 0.99$, consistent with measurements by {\it XMM-Newton} ($\Gamma = 1.75$; Pounds et al. 2001) and with \sax ($\Gamma = 1.58 {+0.09 \atop -0.08}$), although the latter data showed in addition a cut-off at $E_C = 67 {+30 \atop -20} \keV$ \cite{SAXSy1}.

{\bf S5 0716+714}: Results on {\it INTEGRAL} observations of this BL Lac object have been published by Pian et al. 2005. Using the OSA reduction version 4.0 they achieved a marginal detection of $4.5 \sigma$ in the $30 - 60 \keV$ band. Using more data and the current version of the software the source is not detectable, but the strong variability of the source might have resulted in a low average flux. 

{\bf 3C 279}: This blazar ($z = 0.5362$) led to a weak detection with JEM-X and to none with SPI. The spectrum, model shown in Fig.~\ref{fig:3C279}, is dominated by the ISGRI statistics and is well represented by a single powerlaw with photon index $\Gamma = 1.3 {+0.7 \atop -0.5}$ ($\chi^2_\nu = 0.98$), consistent with {\it BeppoSAX} ($\Gamma = 1.66 \pm 0.05$, Ballo et al. 2002) and {\it CGRO}/OSSE observations ($\Gamma = 2.1 {+0.4 \atop -0.4}$; McNaron-Brown et al. 1995). The JEM-X data show an excess in the 2 -- 7 keV energy band. This excess might be caused by the transition from the synchrotron to the inverse Compton branch in this energy range, which would lead to a convex spectrum.

Several sources may have been marginally suspected in the {\it INTEGRAL} data. They were either targets of dedicated observations or were detected in some single science windows. We performed the same analysis as for the other sources presented here, but found only spurious source candidates which result from image reconstruction problems like ghosts, mask patterns, etc. 
Among those sources are {\bf MRK~231}, \mbox{\bf ESO~33--2}, \mbox{\bf PKS~0637--75},
{\bf MCG~--05--23--16}, \mbox{\bf QSO~1028+313}, \mbox{\bf NGC~4736},
{\bf NGC~4418}, \mbox{\bf 3C~353}, and \mbox{\bf QSO~1730--130}$^3$. 

\section{Distribution in Space and Sample Completeness}

In order to explore the space distribution of hard X-ray selected objects, one generally needs to consider a statistically complete flux-limited sample. However, this is difficult to achieve with a data set comprised of various lines of sight and widely differing exposures. In addition, the telescope effective area can be highly position dependent within the coded aperture field of view. Thus, we opt to leave the compilation of a log-N log-S distribution, and perhaps a hard X-ray AGN luminosity distribution to a future paper. 

Alternatively, the $V_e/V_a$-test is a simple method developed by Avni \& Bahcall (1980) based on the $V/V_{max}$ test of Schmidt (1968). $V_e$ stands for the volume, which is enclosed by the object, and $V_a$ is the accessible volume, in which the object could have been found (e.g. due to a flux limit of a survey). Avni \& Bahcall showed that different survey areas with different flux limits in various energy bands can be combined by the $V_e/V_a$-test. In the case of no evolution 
$\langle V_e/V_a \rangle = 0.5$ is expected and the error $\sigma_m(n)$ for a given mean value $\langle m \rangle = \langle V_e/V_a \rangle$ based on $n$ objects is 
\begin{equation}
\sigma_m(n) = \sqrt{\frac{1/3 - \langle m \rangle + \langle m \rangle^2}{n}}
\end{equation}

\begin{figure}
\plotone{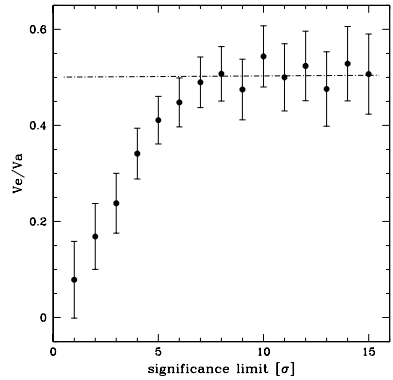}
\caption[]{Result of the $V_e/V_a$-test assuming different completeness limits in ISGRI significance. The dashed dotted line shows the average $\langle V_e/V_a \rangle$ value for the objects with significance $\ge 7\sigma$.}      
\label{fig:VeVa}
\end{figure}

This evolutionary test is applicable only to samples with a well-defined flux limit down to which all objects have been found. It can therefore also be used to test the completeness of a sample. 
We performed a series of $V_e/V_a$-tests to the {\it INTEGRAL} AGN
sample, assuming completeness limits in the range of $1\sigma$ up to
$15\sigma$ ISGRI 20 -- 40 keV significance, adding 5\% systematic
error to the flux uncertainty as described in Section~\ref{data}. For a significance limit
below the true completeness limit of the sample one expects the
$V_e/V_a$-tests to derive a value $\langle V_e/V_a \rangle < \langle
V_e/V_a \rangle_{true}$, where $\langle V_e/V_a \rangle_{true}$ is the
true test result for a complete sample. Above the completeness limit
the $\langle V_e/V_a \rangle$ values should be distributed around
$\langle V_e/V_a \rangle_{true}$ within the statistical
uncertainties. The results of the tests are shown in
Figure~\ref{fig:VeVa}. It appears that the sample becomes complete at
a significance level around $> 6\sigma$, including 31 AGN. The average value
above this significance is  $\langle V_e/V_a \rangle = 0.505 \pm
0.022$. This value is consistent with the expected value of 0.5 which
reflects no evolution and an even distribution in the local
universe. If we do not apply the 5\% systematic error to the data, the
average value is lower ($\langle V_e/V_a \rangle = 0.428 \pm
0.018$), indicating that a systematic error is indeed apparent in the
ISGRI data.

\section{Discussion}

The typical {\it INTEGRAL} spectrum can be described by a simple
powerlaw model with average photon indices ranging from $\Gamma = 2.0$ for
obscured AGN to $\Gamma = 2.1$ for unabsorbed sources (Tab.~\ref{averagespectra}). The simple model does not give reasonable results in cases of high signal-to-noise, where an appropriate fit requires additional features such as a cut-off and a reflection component \citep{Soldi,ngc4151}. 
The results presented here show slightly steeper spectra than previous
investigations of AGN in comparable energy ranges. The same trend is
seen in the comparison of Crab observations, where the {\it
  INTEGRAL}/ISGRI spectra also appear to be slightly steeper than in
previous observations \cite{Kirsch05}, and in comparison of {\it RXTE} and {\it
  INTEGRAL}/ISGRI spectra of Cen~A \cite{Rothschild}.

Zdziarski et al. (1995) did a study of Seyfert galaxies observed by
{\it Ginga} and {\it CGRO}/OSSE, with a similar redshift distribution
(${\bar z} = 0.022$) as the Seyferts detected by {\it INTEGRAL}
(${\bar z} = 0.020$). The same trend in spectral slope from $\Gamma =
1.92 {+0.03 \atop -0.03}$ for radio-quiet Seyfert~1 down to $\Gamma =
1.67 {+0.31 \atop -0.49}$ for Seyfert~2 was detectable in the sample,
and a potential cut-off appeared at energies $E_C > 250 \keV$. The
average of all Seyfert spectra ($\Gamma = 1.9$) appears to be similar
to the one in our sample ($\Gamma = 1.96 \pm 0.01$).

Gondek et al. (1996) confirmed that Seyfert~1 galaxies show a continuum with photon index $\Gamma = 1.9 {+0.1 \atop -0.1}$ with high energy cut-off $E_C > 500 \keV$ and moderate reflection component ($R = 0.76 {+0.15 \atop -0.15}$), based on {\it EXOSAT}, {\it Ginga}, {\it HEAO-1}, and {\it CGRO}/OSSE data.
More recent studies by Perola et al. (2002) on \sax data of 9 Seyfert 1 galaxies confirmed this picture. 
Malizia et al. (2003) used the {\it BeppoSAX} data (3 -- 200 keV) of
14 Seyfert~1 and 17 Seyfert~2 to study average spectral
properties. Also this work finds Seyfert~1 galaxies to be softer
($\Gamma \sim 1.9$) than Seyfert~2 ($\Gamma \sim 1.75$), which also
show stronger absorption. Both types show iron lines with an
equivalent width of 120 eV and a Compton reflection component of $R \sim
0.6 - 1$, but the energy cut-off of the type 1 objects appears to be
at higher energies (200 keV compared to $E_C \sim 130 \, \rm keV$ for the Seyfert~2).

Deluit \& Courvoisier (2003) used for their analysis only the {\it BeppoSAX}/PDS spectra of 14 Seyfert~1, 9 Seyfert~1.5, and 22
Seyfert~2. This study lead to a somewhat different result. Contrary to other studies Seyfert~2 objects show a steeper spectrum ($\Gamma = 2.00 {+0.05 \atop -045}$) than Seyfert~1 ($\Gamma = 1.89 {+0.28 \atop -0.46}$), but with the large error on the results, the difference is not significant. A cut-off does not seem to be required by the \sax data. A marginally significant cut-off is detected in the Seyfert~1 spectra with $E_C = 221 {+\infty \atop -158} \keV$. The average spectra show moderate reflection for Seyfert~1/2 objects ($R = 0.48$ and $R = 0.39$, respectively), while the Seyfert~1.5 appear to have $R = 2.33$ \cite{deluit}. It has to be kept in mind that the results in this study are based on {\it BeppoSAX}/PDS spectra alone, without taking into account the valuable information from the lower energy LECS/MECS data, which could have significantly constrained the shape of the underlying continuum.

The {\it INTEGRAL} AGN sample confirms the previously found trend with
harder spectra for the obscured objects. This can be roughly
understood by the argument that in order to overcome a substantial
absorption, obscured objects have to have hard X-ray spectra to be detectable. 
Comparing the ratio $X$ of obscured ($N_{\rm H} > 10^{22} \rm \, cm^{-2}$) to
unobscured AGN we find in the {\it INTEGRAL} data that $X = 1.7 \pm
0.4$. Excluding the targets of observations, leads to $X = 1.6 \pm
0.5$. The ratios change slightly when taking into account only those
objects which belong to the complete sample with an ISGRI significance
of $7 \sigma$ or higher (Tab.~\ref{catalog}). This sub-sample includes
32~AGN, with 18 obscured and 10 unobscured objects (absorption
information is missing for the remaining four objects).
Using only the complete sample changes the ratio 
to $X = 1.8 \pm 0.5$ and $X = 1.9 \pm 0.6$ (targets excluded),
respectively. 
Splitting this result up into objects near the Galactic
plane ($|b| < 20^\circ$) and off the plane shows for all objects a
ratio of $X = 3.3 \pm 1.1$ and $X = 1.1 \pm 0.5$, respectively. This
trend shows that the harder spectra of those objects, where the absorption in the line of sight through the Galaxy is low compared to the intrinsic absorption, are more likely to shine through the Galactic plane.

One would expect a higher ratio in order to explain the unresolved
X-ray background in this energy range (e.g. Worsley et al. 2005). But
it has to be taken into account that our sample only represents the high flux and 
low luminosity sources ($\langle \log L_X \rangle = 43.3$) in the local
universe ($\langle z \rangle = 0.020$). Gilli et al. (1999) showed that
the ratio between type 2 and all QSOs
should be at most 2 in the local universe. The {\it BeppoSAX} HELLAS
survey showed that the fraction of unobscured AGN increases with the
flux of the objects \cite{HELLAS}. Evidence that the fraction
of obscured AGN depends on the redshift was found by Ueda et
al. (2004) when studying the hard X-ray luminosity function. They find
that the density evolution of AGN depends on the luminosity which will
lead to an increase of the type 2 fraction with redshift. In a recent study La Franca et al. (2005) find that both effects combined (the fraction of absorbed AGN decreases with the intrinsic X-ray luminosity, and increases with the redshift) can be explained by a luminosity-dependent density evolution model. They further show that the luminosity function of AGN, like those presented here, peaks at $z \sim 0.7$ while high luminosity AGN peak at $z \sim 2$. Consistent with our study, La Franca et al. also find a ratio of $X = 2.1$ at $L_X = 10^{42.5} \rm \, erg \, s^{-1}$ in the $2 - 10 \, \rm keV$ energy band.
Unified models predict, depending on the applied model, a ratio of $X
= 1.5 - 2.0$ for high flux low redshift AGN \cite{AGNunification}, while 
a recent study of {\it Chandra} data showed that $X = 8.5 \pm 6.3$ for
low redshift, low luminosity AGN \cite{Barger}, indicating a larger
dominance of obscured objects. These investigations have been focused
so far on the X-rays below 10 keV, and {\it INTEGRAL} can add
substantial information to the nature of bright AGNs in the local
universe. Considering the expected composition of the hard X-ray background, it is remarkable that our sample includes only 4 Compton thick AGN. If Compton thick AGN are dominant in the hard X-rays, they would have to be fainter than the objects detected by {\it INTEGRAL} so far.   

The energy range in the 15 -- 200 keV is now also accessible through
the BAT instrument aboard {\it Swift} \cite{Swift}. A study by
Markwardt et al. (2005) of observations of the extragalactic sky indicates a ratio of $X = 2$
between obscured and unobscured AGN, fully consistent with the
results from the {\it INTEGRAL} AGN sample. Although the {\it Swift}/BAT survey
covers different areas of the sky, the energy band is similar to the
{\it INTEGRAL}/ISGRI range and the type of AGNs detectable should be
the same.

The sample presented here is still too small to constrain the
ratio of obscured to unobscured sources, but it might indicate that the unified model predicts this value of $X \simeq 2$ correctly, differing from earlier claims \citep{AGNunification,LaFranca}.

\section{Conclusions}

The AGN sample derived from the {\it INTEGRAL} public data archive
comprises 42 low luminosity and low redshift ($\langle \log L_X \rangle = 43.3$, $\langle z \rangle = 0.020 \pm 0.004$, excluding the blazars) objects including 5~blazars in the hard X-ray domain. Only one galaxy cluster is also detected, but no star-burst galaxy has been as yet.
The results from the JEM-X2, ISGRI, and SPI data show that within the
statistical limitations of the data the high-energy continua of AGN
can be described by a simple powerlaw with photon index $\Gamma = 2.11
\pm 0.05$ (Seyfert 1),  $\Gamma = 2.10 \pm 0.02$ (Seyfert 1.5), and
$\Gamma = 1.95 \pm 0.01$ (Seyfert 2). A similar trend is seen when
dividing the Seyfert galaxies according to their intrinsic absorption into
unabsorbed ($N_H < 10^{22} \rm \, cm^{-2}$; $\Gamma = 2.08 \pm 0.02$)
and absorbed sources ($\Gamma = 1.98 \pm 0.01$). Blazars naturally
show strongly variable spectra, but the average spectral slope is of
the same order ($\Gamma = 2.07 \pm 0.10$). It has to be taken into
account that the simple model does not represent the data as soon as a
sufficient signal-to-noise is apparent, like in the Circinus galaxy and Cen~A \citep{Soldi,Rothschild} or NGC~4151 \cite{ngc4151}. But the average spectra indicate that a cut-off in most cases, if apparent, appears at energies $E_C \gg 200 \keV$.

The AGN sample presented here is complete down to a significance limit of $\sim 7 \sigma$. This results in a complete sample of 32~AGN.
Within this complete sample the ratio between obscured and unobscured
AGN is $X = 1.8 \pm 0.5$, consistent with the unified model for AGN
and with recent results from the {\it Swift}/BAT survey. Only 4 Compton thick AGN are found in the whole and only 3 in the complete sample.

Sazonov et al. (2005) report five additional AGN in
recent observations, not detectable in the data presented here. Additionally one AGN seen by OSSE, the blazar \mbox{3C~454.3} (McNaron-Brown et al. 1995), has recently been detected by {\it INTEGRAL} during an outburst \cite{3C454INTEGRAL}.
With the ongoing observations of the {\it INTEGRAL} and {\it Swift} mission, the
number of detectable AGNs will increase further, especially with a
number of already performed and scheduled observations far off the
Galactic plane. This will enable us to solve the mystery of the hard
X-ray background above 10 keV in the near future.

\begin{acknowledgements}
We like to thank the anonymous referee for the comments which helped to improve the paper. This research has made use of the NASA/IPAC Extragalactic Database (NED) which is operated by the Jet Propulsion Laboratory, of data obtained from the High Energy Astrophysics Science Archive Research Center (HEASARC), provided by NASA's Goddard Space Flight Center, and of the SIMBAD Astronomical Database which is operated by the Centre de Donn\'ees astronomiques de Strasbourg. This research has made use of the Tartarus (Version 3.1) database, created by Paul O'Neill and Kirpal Nandra at Imperial College London, and Jane Turner at NASA/GSFC. Tartarus is supported by funding from PPARC, and NASA grants NAG5-7385 and NAG5-7067. 
\end{acknowledgements}

%
%
%

%

%
%
\begin{deluxetable}{lclrrrcrcccc}
\tabletypesize{\scriptsize}
\tablecaption{{\it INTEGRAL} AGN catalog\label{catalog}}
\tablehead{
\colhead{Name} & \colhead{Type} & \colhead{z} & \colhead{R.A.} & \colhead{DEC} &
\colhead{exp.\tablenotemark{a}} & \colhead{target} & \colhead{ISGRI} &
\colhead{SPI} & \colhead{JEM-X} & \colhead{OSSE\tablenotemark{b}} \\
\colhead{} & \colhead{} & \colhead{} & \colhead{} & \colhead{} & \colhead{[ks]} & \colhead{} & \colhead{$[\sigma]$} & \colhead{$[\sigma]$} & \colhead{$[\sigma]$}
}
\startdata
NGC 788 & Sy 1/2 & 0.0136 &   02 01 06 & --06 46 56& 311 & -- & 10.1 &
-- & -- & --\\ 
NGC 1068 & Sy 2 & 0.003793 &   02 42 41 & --00 00 48& 311 & -- & 4.4 &
-- & -- & --\\ 
NGC 1275 & Sy 2  & 0.017559  & 03 19 48 & +41 30 42 & 264 & -- & 19.1 & -- & 78 & $1.8 \pm 0.5$\\
3C 111   & Sy 1  & 0.048500  & 04 18 21 & +38 01 36 & 67 & --& 9.5 & 0.1 & -- & $2.7 \pm 0.5$\\
MCG +8--11--11 &Sy 1.5& 0.020484 & 05 54 54&+46 26 22 & 21 & -- & 5.5 & -- & -- & $3.7 \pm 0.3$\\
MRK 3    & Sy 2 & 0.013509 & 06 15 36 & +74 02 15 & 472 & -- & 13.3 & 2.9 & -- & $1.8 \pm 0.4$\\
MRK 6    & Sy 1.5 & 0.018813 & 06 52 12 & +74 25 37 & 482 & -- & 8.8 & 1.1 & -- & --\\
NGC 4051 & Sy 1.5  & 0.002336 & 12 03 10 & +44 31 53 & 443 & -- & 8.8   & 3.0 & -- & --\\
NGC 4151 & Sy 1.5  & 0.003320 & 12 10 33 & +39 24 21 & 483 & X & 163.3 & 27.2& 2089 & $24.5 \pm 0.4$\\
NGC 4253 & Sy 1.5  & 0.012929 & 12 18 27 & +29 48 46 & 715 & -- &  4.3 & 3.1 & -- & -- \\
NGC 4388 & Sy 2    & 0.008419 & 12 25 47 & +12 39 44 & 215 & X & 37.9 & 4.0 & 19 & $9.1 \pm 0.4$\\
NGC 4395 & Sy 1.8  & 0.001064  & 12 25 49 & +33 32 48 & 739 & -- & 10.0 & -- & -- & -- \\
NGC 4507 & Sy 2    & 0.011801 & 12 35 37 & --39 54 33 & 152 & -- & 15.8 & 3.6 & n/a & $5.2 \pm 0.6$\\
NGC 4593 & Sy 1    & 0.009000 & 12 39 39 & --05 20 39 & 723 & -- & 20.9 & 3.8 & 3 & $< 2.9$\\
Coma Cluster & GClstr & 0.023100 & 12 59 48 & +27 58 48 & 516 & X & 6.6 & 1.4 & 8.9 & --\\
NGC 4945 & Sy 2 & 0.001878   & 13 05 27 & --49 28 06 & 276 & X & 33.4 & 6.7 & -- & $10.9 \pm 0.7$\\
ESO 323--G077 & Sy 2  & 0.015014  & 13 06 26 & --40 24 53 & 761 & -- & 9.9 & 1.3 & -- & --\\
NGC 5033 & Sy 1.9  & 0.002919 & 13 13 28 & +36 35 38 & 377 & -- & 2.9 & 1.2 & -- & -- \\
Cen A    & Sy 2    & 0.001830 & 13 25 28 & --43 01 09 & 532 & X & 166.1 &24.8 & 1479 & $42.5 \pm 0.4$\\
MCG--06--30--015 & Sy 1 & 0.007749  &13 35 54 & --34 17 43 & 567 & X & 2.9 & 2.2 & 9 & $4.1 \pm 0.6$\\
4U 1344--60 & ? & 0.043  & 13 47 25 & --60 38 36 & 603 & -- & 40.9 & 5.9 & -- & --\\ 
IC 4329A & Sy 1.2  & 0.016054  & 13 49 19 & --30 18 36 & 440 & X & 43.5 & 6.2 & 1 & $7.5 \pm 0.4$\\
Circinus gal.& Sy 2 & 0.001448 & 14 13 10& --65 20 21 & 589 & -- & 79.5 & 8.6 & 2 & --\\
NGC 5506 & Sy 1.9  & 0.006181  & 14 13 15 & --03 12 27 & 101 & X & 19.6 & n/a & 39 & $5.0 \pm 0.6$\\ 
PG 1416-129 & Sy 1 & 0.129280 & 14 19 04 &--13 10 44 & 117 & --  & 12.0  &  1.1 & n/a & --\\
IC 4518  & Sy 2 &  0.015728 & 14 57 43 & --43 07 54 & 338 & -- & 4.8 & 0.5 & -- & --\\
NGC 6221 & Sy 1/2 & 0.004977 & 16 52 46 & --59 13 07 & 523 & -- & 7.6 & 7.0 & -- & --\\
NGC 6300 & Sy 2   & 0.003699 & 17 17 00 & --62 49 14 & 173 & -- &  9.8 & -- & 3 & --\\
GRS 1734-292 & Sy 1 & 0.021400 & 17 37 24 & --29 10 48 & 3332 & -- & 43.8 & 16.9 & 15.2 & --\\
IGR J18027-1455&Sy 1& 0.035000 & 18 02 47 & --14 54 55 & 942 & -- & 12.1 & 5.8 & 14 & --\\
ESO 103-G35& Sy 2  & 0.013286 & 18 38 20 & --65 25 39 & 36 & -- & 12.2 & 1.5 & -- & --\\
1H 1934-063& Sy 1 & 0.010587 & 19 37 33 & --06 13 05 & 684 & -- & 6.6 & 4.4 & -- & --\\
NGC 6814 & Sy 1.5 & 0.005214 & 19 42 41 & --10 19 25 & 488 & -- & 13.9 & 1.7 & -- & $2.8 \pm 0.5$\\
Cygnus A & Sy 2 & 0.056075 & 19 59 28 & +40 44 02 & 1376 & -- & 7.3 & 2.5 & -- & --\\
MRK 509  & Sy 1 & 0.034397   & 20 44 10 & --10 43 25 & 73 & X & 10.1 & 2.6 & 6 & $3.6 \pm 0.6$\\
IGR J21247+5058&radio gal.?& 0.3\tablenotemark{c} & 21 24 39&+50 58 26& 213 & -- & 30.8 & 5.2 & -- & --\\ 
MR 2251-178& Sy 1 & 0.063980 & 22 54 06 & --17 34 55 & 489 & X & 35.9 & 1.9 & -- & --\\
MCG -02-58-022 & Sy 1.5 & 0.046860 & 23 04 44 & --08 41 09 & 489 & -- & 5.9  & 1.6 & -- & $3.3 \pm 0.8$\\[0.5cm]

S5 0716+714 & BL Lac & 0.3\tablenotemark{c}& 07 21 53 & +71 20 36 & 482 & X & 0.7 & -- & -- & --\\
S5 0836+710 & FSRQ  & 2.172  & 08 41 24 & +70 53 42 & 391 & -- & 8.5 & 0.6 & -- & --\\
3C 273   & Blazar & 0.15834  & 12 29 07 & +02 03 09 & 655 & X & 35.6 & 6.0 & 32 & $10.6 \pm 0.3$\\
3C 279   & Blazar &  0.53620 & 12 56 11 & +05 47 22 & 497 & X & 6.8 & -- & 3 & $3.3 \pm 0.5$\\
PKS 1830--211&Blazar& 2.507 & 18 33 40 & --21 03 40 & 1069 & -- & 16.1 & 1.4 & -- & --\\
\enddata

\tablenotetext{a}{ISGRI exposure time}
\tablenotetext{b}{flux in $50 - 150 \rm \, keV$ in units of $10^{-4} \rm \, ph \, cm^{-2} \, s^{-1}$}
\tablenotetext{c}{tentative redshift}

\end{deluxetable}
\begin{table}
\caption[]{{\it CGRO}/OSSE detected AGN not seen by {\it INTEGRAL}}
\begin{tabular}{lcr}
\tableline\tableline
Name & Type & exposure$^a$\\
     &      & [ks]      \\
\hline
3C 120 & Sy1 & 3\\
3C 390.3 & Sy1 & --\\
3C 454.3 & blazar & 2\\
CTA 102  & blazar & 6\\
ESO 141-G55 & Sy 1 & 23\\
H 1517+656 & BL Lac& 2\\
III Zw2 & Sy1/2 & 79\\
MRK 279 & Sy1.5 & 13\\
MRK 841 & Sy1.5 & --\\
M82 & Starburst & 84\\
NGC 253 & Starburst & --\\
NGC 2110 & Sy2 & --\\
NGC 3227 & Sy1.5 & --\\
NGC 3783 & Sy1 & --\\
NGC 526A & Sy1.5 & --\\
NGC 5548 & Sy1.5 & --\\
NGC 7172 & Sy2 & --\\
NGC 7213 & Sy1.5 & --\\
NGC 7314 & Sy1.9 & 489\\
NGC 7469 & Sy1.2 & 2\\
NGC 7582 & Sy2 & --\\
PKS 0528+134 & blazar & 1040\\
PKS 2155-304 & BL Lac & --\\
QSO 1028+313 & Sy1.5 & --\\

\tableline
\end{tabular}
\label{OSSEnonINTEGRAL}

$^{a}$ ISGRI exposure time\\
\end{table}
\begin{deluxetable}{lrrrcrc}
\tabletypesize{\scriptsize}
\tablecaption{{\it INTEGRAL} AGN fit results\label{fitresults}}
\tablehead{
\colhead{Name} & \colhead{ISGRI\tablenotemark{a}} &
\colhead{ISGRI\tablenotemark{a}} & \colhead{JEM-X\tablenotemark{a}} &
\colhead{$\log L_X$\tablenotemark{b}} & \colhead{$N_{\rm H}$} &
\colhead{$\Gamma$} \\ \colhead{} & \colhead{$f_{\rm 20 - 40 \, keV}$}
& \colhead{$f_{\rm 40 - 100 \, keV}$} &  \colhead{$f_{\rm 2 - 10 \,
    keV}$} & \colhead{[20 -- 100 keV]} & \colhead{$[10^{22} \rm \, cm^{-2}]$} & \colhead{}  
}
\startdata
NGC 788  & $2.98 \pm 0.24$ & $5.06 \pm 0.41$ & --  & 43.52 & $<
0.02$\tablenotemark{e}& $1.69 {+0.31 \atop -0.27}$\\
NGC 1068 & $0.93 \pm 0.27$ & $1.66 \pm 0.48$ &     --          & 41.92
& $> 150$\tablenotemark{n}& $1.71 {+0.82 \atop -0.80}$\\
NGC 1275 & $1.89 \pm 0.21$ & $2.29 \pm 0.25$ & $4.34 \pm 0.03$ & 43.47
& 3.75\tablenotemark{d} & $2.12 {+0.60 \atop -0.43}$\\
3C 111   & $6.27 \pm 0.57$ & $8.10 \pm 0.74$ & -- & 44.90 & 0.634\tablenotemark{e} & $2.03 {+0.45 \atop -0.37}$\\
MCG +8-11-11 & $6.07 \pm 0.97$ & $5.80 \pm 0.92$ & -- & 44.06 &
$< 0.02$\tablenotemark{f} & $2.41 {+1.24 \atop -0.70}$ \\
MRK 3    & $3.65 \pm 0.39$ & $4.62 \pm 0.64$ & -- & 43.53 & 110\tablenotemark{f} & $2.05 {+0.27 \atop -0.23}$\\ 
MRK 6    & $2.01 \pm 0.20$ & $2.68 \pm 0.26$ & -- & 43.57 & 10\tablenotemark{f} & $2.01 {+0.42 \atop -0.35}$\\
NGC 4051 & $1.80 \pm 0.20$ & $1.48 \pm 0.17$ & -- & 43.60 & $<
0.01$\tablenotemark{f} & $2.62 {+0.81 \atop -0.53}$\\
NGC 4151 & $26.13 \pm 0.16$ & $31.97 \pm 0.20$ & $20.89 \pm 0.01$ &
43.15 & 4.9\tablenotemark{g} & $2.10 \pm 0.02$ \\
NGC 4253 & $0.93 \pm 0.22$ & $1.38 \pm 0.32$ & --& 42.93 &
0.8\tablenotemark{f} & $1.85 {+0.78 \atop -0.60}$\\
NGC 4388 & $ 9.54 \pm 0.25$ & $13.08 \pm 0.35$ & $4.63 \pm 0.25$ & 43.55 & 27\tablenotemark{h} & $1.92 {+0.09 \atop -0.09}$ \\
NGC 4395 & $0.56 \pm 0.22$ & $1.20 \pm 0.47$ & -- & 40.64 & 0.15\tablenotemark{e}& $1.31 {+0.85 \atop -0.71}$ \\
NGC 4507 & $6.46 \pm 0.36$ & $ 9.11 \pm 0.50$ & -- & 43.68 & 29\tablenotemark{f} &$1.92 {+0.35 \atop -0.31}$ \\
NGC 4593 & $3.31 \pm 0.16$ & $3.39 \pm 0.16$ & $2.90 \pm 0.89$ & 43.08 & 0.02\tablenotemark{f}& $1.02 {+0.63 \atop -0.59}$ \\
Coma Cluster & $1.09 \pm 0.12$ & $1.00 \pm 0.15$ & $8.82 \pm 1.27$ & 43.40 & $<0.01$\tablenotemark{d} &$2.65 {+0.18 \atop -0.18}$ \\
NGC 4945 & $ 9.85 \pm 0.23$ & $15.78 \pm 0.37$ & -- & 42.30 &
400\tablenotemark{f} & $1.93 {+0.09 \atop -0.08}$\\
ESO 323-G077 & $1.20 \pm 0.19$ & $2.03 \pm 0.33$ & -- & 43.21 &
55\tablenotemark{i} & $1.72 {+0.51 \atop -0.33}$\\
NGC 5033 & $1.06 \pm 0.24$ & $0.84 \pm 0.19$ & -- & 41.55 &
2.9\tablenotemark{f} & $2.7 {+7.3 \atop -1.2}$ \\
Cen A    & $32.28 \pm 0.17$ & $43.79 \pm 0.23$ & $29.58 \pm 0.02$ &
42.75 & 12.5\tablenotemark{d} &$1.94 {+0.02 \atop -0.02}$ \\
MCG-06-30-015 & $2.98 \pm 0.19$ & $3.50 \pm 0.22$ & $4.46 \pm 0.51$ & 42.93 &17.7\tablenotemark{f} & $2.04 {+0.10 \atop -0.11}$ \\
4U 1344-60 & $2.83 \pm 0.15$ & $3.73 \pm 0.20$ & -- & 44.36 &
2.19\tablenotemark{f} & $1.65 {+0.02 \atop -0.03}$\\
IC 4329A & $8.19 \pm 0.17$ & $10.71 \pm 0.23$ & $0.98 \pm 1.21$ &
44.04 & 0.42\tablenotemark{f} &$2.01 {+0.09 \atop -0.08}$\\
Circinus gal.& $10.73 \pm 0.18$ & $ 9.39 \pm 0.15$ & $1.29 \pm 0.73$ &
41.97 & 360\tablenotemark{f} &$2.74 {+0.09 \atop -0.09}$\\
NGC 5506 & $4.21 \pm 0.33$ & $ 3.69 \pm 0.30$ & $8.87 \pm 0.23$ &
42.83 & 3.4\tablenotemark{f} & $2.15 {+0.09 \atop -0.08}$\\ 
PG 1416-129 & $5.43 \pm 0.64$ & $ 8.62 \pm 1.01$ & -- & 45.78 &0.09\tablenotemark{e} & $1.77
{+0.51 \atop -0.41}$\\
IC 4518 & $0.49 \pm 0.32$ & $1.04 \pm 0.68$ & -- & 42.92 & ? & $1.48
{+3.47 \atop -1.57}$\\
NGC 6221 & $1.32 \pm 0.20$ & $1.77 \pm 0.27$ & -- & 42.39 &
1\tablenotemark{k} & $1.82 {+0.35 \atop -0.28}$ \\
NGC 6300 & $3.91 \pm 0.37$ & $3.72 \pm 0.35$ & $4.66 \pm 1.69$ & 42.36
& 22\tablenotemark{l} & $2.42 {+0.54 \atop -0.41}$\\
GRS 1734-292 & $4.03 \pm 0.09$ & $3.13 \pm 0.07$ & $68.4 \pm 4.5$ &
43.88 & 3.7\tablenotemark{d} & $2.67 {+0.12 \atop -0.11}$ \\ 
IGR J18027-1455& $2.03 \pm 0.16$ & $1.66 \pm 0.13$ & $4.38 \pm 0.31$ & 44.03 &19.0\tablenotemark{d} & $2.60 {+0.56 \atop -0.42}$\\
ESO 103-G35 & $2.97 \pm 0.66$ & $ 5.27 \pm 1.17$ & -- & 43.51 & 13 -- 16\tablenotemark{e} &$1.64
{+0.97 \atop -0.70}$\\
1H 1934-063& $0.48 \pm 0.25$ & $0.83 \pm 0.44$ & -- & 42.51 & ? & $2.92 {+0.81 \atop -0.63}$ \\
NGC 6814 & $2.92 \pm 0.23$ & $2.64 \pm 0.21$ & -- & 42.52 & $< 0.05$\tablenotemark{e} &$2.48 {+0.49 \atop -0.36}$ \\
Cygnus A & $3.24 \pm 0.14$ & $3.55 \pm 0.15$ & -- & 44.71 & 20\tablenotemark{m} &$2.24
{+0.19 \atop -0.17}$ \\
MRK 509  & $4.66 \pm 0.47$ & $4.96 \pm 0.50$ & $4.56 \pm 0.78$ & 44.42
& $< 0.01$\tablenotemark{f} &$1.73 {+0.15 \atop -0.16}$\\
IGR J21247+5058& $4.15 \pm 0.27$ & $6.85 \pm 0.45$ & -- & 44.00$^c$ &
? & $1.73 {+0.24 \atop -0.22}$\\ 
MR 2251-178 & $1.20 \pm 0.17$ & $1.27 \pm 0.18$ & -- & 44.40 & 0.02 --
0.19\tablenotemark{e} &$2.31
{+1.21 \atop -0.68}$\\ 
MCG -02-58-022 & $1.20 \pm 0.28$ & $1.75 \pm 0.41$ & -- & 44.18 & $<0.01 -
0.08$\tablenotemark{e} &
$1.71 {+0.89 \atop -0.63}$\\[0.5cm]

S5 0716+714 & $0.14 \pm 0.11$ & $0.71 \pm 0.59$ & -- & 45.21$^c$ & $<
0.01$\tablenotemark{e} & -- \\ 
S5 0836+710 & $1.73 \pm 0.28$ & $3.02 \pm 0.50$ & -- & 47.87 &
$0.11$\tablenotemark{e} &  $1.65 {+0.46 \atop -0.39}$\\ 
3C 273   & $5.50 \pm 0.15$ & $6.29 \pm 0.17$ & $9.21 \pm 0.29$ & 45.92
& $0.5$\tablenotemark{d} &$2.17{+0.12 \atop -0.12}$\\
3C 279   & $0.82 \pm 0.24$ & $1.79 \pm 0.52$ & $6.05 \pm 2.36$ & 46.37
& $0.02 - 0.13$\tablenotemark{e} &$1.3 {+0.7 \atop -0.5}$\\
PKS 1830--211& $2.07 \pm 0.14$ & $2.82 \pm 0.19$ & -- & 48.09 & $<0.01
- 0.7$\tablenotemark{e} & $1.96 {+0.27 \atop -0.24}$
\enddata
\tablenotetext{a}{flux in $10^{-11} \, \rm erg \, cm^{-2} \, s^{-1}$}
\tablenotetext{b}{ luminosity in $\rm erg \, s^{-1}$ for $H_0 = 70$ and $\Lambda_0 = 0.73$}
\tablenotetext{c}{tentative redshift}
\tablenotetext{d}{this work, $^{\rm e}$Tartarus database, $^{\rm
    f}$Lutz et al. 2004, $^{\rm g}$Beckmann et al. 2005, $^{\rm
    h}$Beckmann et al. 2004, $^{\rm i}$Sazonov \& Revnivtsev 2004,
  $^{\rm k}$Levenson, Weaver \& Heckman 2001, $^{\rm l}$Matsumoto et
  al. 2004, $^{\rm m}$Young et al. 2002, $^{\rm n}$Matt et al. 1997a}  
\end{deluxetable}
\begin{table}
\caption[]{Cut-off power law fits}
\begin{tabular}{lcc}
\tableline\tableline
source & $\Gamma$ & $E_C$\\ 
 & powerlaw & [keV]\\ 
\hline
NGC~4507 & $1.0 {+1.7 \atop -0.8}$ & $55 {+123 \atop -11}$\\
NGC~4593 & $1.02 {+0.63 \atop -0.59}$ & $35 {+49 \atop -12}$\\
ESO~323-G077  & $1.48 {+0.65 \atop -1.18}$ & $181 {+319 \atop -131}$\\
IC~4329A & $1.51 {+0.39 \atop -0.40}$ & $104 {+344 \atop -48}$\\
Circinus~Galaxy & $2.23 {+0.48 \atop -0.55}$ & $87  {+\infty \atop -46}$\\
NGC~5506 & $1.81 {+0.25 \atop -0.28}$ & $57  {+139 \atop -27}$\\

\tableline
\end{tabular}
\label{cutoff}
\end{table}

\begin{table}
\caption[]{Average spectra}
\begin{tabular}{lc}
\tableline\tableline
source type& $\Gamma^a$\\
\hline
unabsorbed Sy & $2.08 \pm 0.02$\\
absorbed Sy & $1.98 \pm 0.01$\\
Seyfert 1   & $2.11 \pm 0.05$ \\
Seyfert 1.5 & $2.10 \pm 0.02$ \\
Seyfert 2   & $1.95 \pm 0.01$ \\
Blazar      & $2.07 \pm 0.10$ \\

\tableline
\end{tabular}
\label{averagespectra}

$^a$ weighted mean of the individual source results\\
\end{table}

\end{document}